\documentclass[preprint,prc,aps,epsfig]{revtex4}
\usepackage{psfig}
\def\beq{\begin{equation}}
\def\enq{\end{equation}}
\def\beqa{\begin{eqnarray}}
\def\enqa{\end{eqnarray}}

\def\GeV{\nobreak\,\mbox{GeV}}

\def\pli{p^\prime}
\def\mli{{M^\prime}^2}

\def\ka{\kappa}
\def\la{\lambda}
\def\ga{\gamma}
\def\Ga{\Gamma}

\def\si{\sigma}

\def\lb{\label}
\def\nn{\nonumber}
\newcommand{\rag}{\rangle}
\newcommand{\lag}{\langle}

\newcommand{\rf}{\ref}

\begin{document}

\title{\sc The structure of $f_0(980)$ from charmed mesons decays}
\author {Ignacio Bediaga$^1$, Fernando S. Navarra$^2$ and Marina Nielsen$^2$}
\affiliation{ $^1$Centro Brasileiro de Pesquisas F\'\i sicas\\
Rua Xavier Sigaud 150, 22290-180 Rio de Janeiro, RJ, Brazil\\[0.1cm]
$^2$Instituto de F\'{\i}sica, Universidade de S\~{a}o Paulo\\
  C.P. 66318,  05315-970 S\~{a}o Paulo, SP, Brazil}
\begin{abstract}
We use the QCD sum rules to evaluate the form factors associated with
the semileptonic decays of $D_s$ and $D$ mesons into $f_0(980)$. 
We consider the $f_0(980)$ meson as a quark-antiquark state
with a mixture of strange and light components. The decay rates are evaluated
in terms of the mixing angle. Using the same form factors to evaluate
nonleptonic decays in the framework of the factorization approximation
we conclude that the importance of the light quarks in $f_0(980)$
is not negligible.
\end{abstract}

\maketitle

Recent experimental data coming from charmed mesons decays
 have opened new possibilities to understand the spectroscopy 
of the controversial light scalar mesons, since they are abundantly 
 produced in these decays \cite{ex}. Actually, with the firm identification
of the scalars $\sigma$ and $\kappa$,
the observed scalar states below $1.5\GeV$
are too numerous  to be accommodated in a single $q\bar{q}$ multiplet.
This proliferation of the scalar mesons is consistent with two nonets, 
one below 1 GeV region  and another one near 1.5 GeV \cite{cloto}. 
In this new scheme the light scalars (the
isoscalars $\si(500),\;f_0(980)$, the isodoublet $\kappa$ and
the isovector $a_0(980)$) would form an SU(3) flavor nonet. In  the
naive quark
model the flavor structure of these scalars would generically be:
\beqa
\si={\cos(\alpha)\over\sqrt{2}}(u\bar{u}+d\bar{d})-\sin(\alpha)s\bar{s}
,\;\;\;\;\;\;&&f_0=\cos(\alpha)s\bar{s}+{\sin(\alpha)\over\sqrt{2}}
(u\bar{u}+d\bar{d}),
\label{f0stru}\\
a_0^0={1\over\sqrt{2}}(u\bar{u}-d\bar{d}),\;\;\;\;\;\;&&a_0^+=u\bar{d},
\;\;\;\;\;\;a_0^-=d\bar{u},
\\
\ka^+=u\bar{s},\;\;\;\;\;\;\ka^0=d\bar{s},\;\;\;\;\;\;&&\bar{\ka}^0=s\bar{d},
\;\;\;\;\;\;\ka^-=s\bar{u},
\enqa
where we have already allowed a mixing between the isoscalars
$s\bar{s}$ and $(u\bar{u}+d\bar{d})$. Although the predominant $s\bar{s}$
nature of the $f_0(980)$ has been supported by the radiative decay
$\phi\to f_0(980)\gamma$, and by some theoretical calculations 
\cite{torn,bev}, there is no  fundamental  theoretical  
reason to expect $\alpha=0$, as we find in the vector meson sector. 
Actually, since instantons are supposed to be as important in the scalar 
sector as they are in the pseudoscalar sector \cite{inst}, we would expect
a mixing in the scalar sector similar to what we have in the pseudoscalar 
sector. In this sense, the measurements of  
$J/\psi\rightarrow f_0(980)\phi$ and $J/\psi\rightarrow f_0(980)\omega$
with similar branching ratios \cite{PDG},
indicating that $f_0(980)$ is not purely an $s\bar{s}$ state, 
can not be taken as a surprise.
In ref.~\cite{cheng}  these $J/\psi$ decays were used to estimate the
 mixing angle in Eq.~(\ref{f0stru}), giving $\alpha=(34\pm6)^0$. A similar 
 mixing angle, $35^0\leq\alpha\leq55^0$ \cite{adn}, 
was found analysing the experimental results $D_s^+\to f_0(980)
\pi^+$ and $D_s^+\to \phi\pi^+$ \cite{E791}.  Using $f_0(980)$ as a 
pure $s\bar{s}$ state ({\it i.e.}, using $\alpha=0$) 
the authors of  ref.~\cite{phi}, 
 could not reproduce the  experimental result of 
  $D_s^+\to f_0(980)\pi^+$/ $D_s^+\to \phi\pi^+$ \cite{E791}. They 
 concluded that there is room for a sizable light quark component in
$f_0(980)$, corresponding to a mixing angle of about $\alpha\sim40^0$.

In this work we propose that experimental results of the 
semileptonic decays $D_s^+\to f_0(980)
\ell^+\nu_\ell$ and $D^+\to f_0(980)\ell^+\nu_\ell$, can be
used to get the minimum bias estimate of the importance of the light quark 
content in the scalar-isoscalar $f_0(980)$, in the quark-antiquark scenario. 
 The hadronic part of the current of the $D_s^+$ decay, can 
 produce the $f_0(980)$ only through  the $s \bar s$ component, while 
 the hadronic part of the $D^+$ decay  can produce the $f_0(980)$ only 
through  $d \bar d$. To observe these decays it would be necessary 
 a high statistics experiment and also a tagging to separate the 
 semi-leptonic $D_s^+$  from the   $D^+$ decays. These two conditions 
could be  satisfied next year by  the CLEO-C experiment \cite{cleo-c}.

In order to estimate theoretically
these semileptonic $D_s^+$ and $D^+$ decays, we use the 
method of QCD  sum rules \cite{svz} which has been
successfully applied to several semileptonic decay processes.  
Since the semileptonic decay is supposed  to occur on
the quark level, the decay rate should depend  crucially on the
direct coupling of the resonances to the quark currents.   
The coupling of the $f_0(980)$ to the scalar current
\beq
j_s=\bar{s}s\cos(\alpha)+(\bar{u}u+\bar{d}d){\sin(\alpha)\over\sqrt{2}},
\label{jf0}
\enq
can be parametrized as
\beq
\langle0|j_s|f_0(p)\rangle=\lambda_{f_0}=m_{f_0}f_{f_0},
\label{coup}
\enq
and can be determined by the QCD sum rule based on the two-point
correlation function
\beq
\Pi(q^2)=i\int d^4x e^{iq.x}\lag0|{\rm T}[ j_s(x) j_s^\dagger(0)]|0\rag.
\enq
This same correlation function was studied in ref.~\cite{faz} in the case 
$\alpha=0$. Following the same procedure and considering a general
mixing angle $\alpha$ we get the sum rule (up to order $m_s$):
\beqa
\lambda_{f_0}^2e^{-m_{f_0}^2/M^2}&=&\cos^2(\alpha)\left[
{3\over8\pi^2}\int_0^{u_0}
du u e^{-u/M^2}\right.\nn\\
&+&\left.
m_se^{-m_s^2/M^2}\left(3\lag\bar{s}s\rag
+{\lag\bar{s}g_s\si.Gs\rag\over M^2}\right)
\right]\nn\\
&+&\sin^2(\alpha)\left({3\over8\pi^2}\int_0^{u_0}du u e^{-u/M^2}\right)\;.
\lb{2f0}
\enqa

\begin{figure} \label{fig1}
\centerline{\psfig{figure=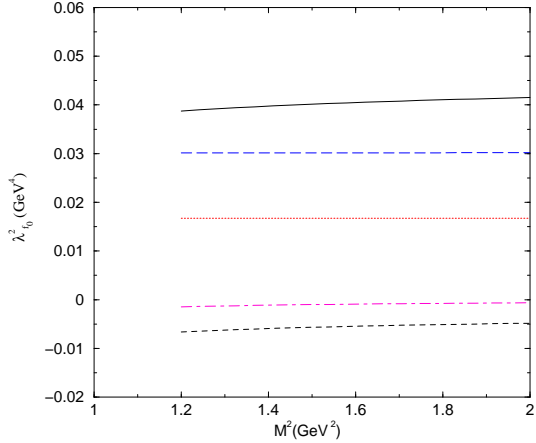,height=60mm,width=70mm,angle=0}}
\caption{Various contributions to the OPE of the coupling $\lambda_{f_0}^2$,
as a function of the Borel parameter $M^2$. The solid line gives the
final result and the long-dashed, dashed, dot-dashed and dotted lines
give the contributions of the first, second, third and fourth  terms
in Eq.~(\ref{2f0}).} 
\end{figure}
In the numerical analysis of the sum rules, the values used for the strange
quark mass and condensates are: $m_s=0.14\,\GeV$, $\lag\bar s s\rag=\,
0.8\lag\bar{q}q\rag$, $\lag\bar{q}q\rag=-(0.23)^3\,\GeV^3$,
$\lag\bar{s}g_s\si.Gs\rag =m_0^2
\lag\bar{s}s\rag$ with $m_0^2=0.8\,\GeV^2$.
We evaluate the 2-point sum rules in  the same stability window found
in \cite{faz}: $1.2\leq M^2\leq2.0\GeV^2$. 
In Fig.~1 we show the different contributions to $\lambda_{f_0}^2$
as a function of the Borel mass
using the continuum threshold $u_{0}=1.6~\GeV^2$ and the mixing angle 
$\alpha=37^0$. We see that $\lambda_{f_0}^2$ is very stable, as a function 
of the Borel mass, in the considered Borel range. The coupling is 
not very sensitive to changes in the values of the continuum
threshold and mixing angle. For 
$u_0=(1.6\pm0.1)\GeV^2$ and $0\leq\alpha\leq37^0$ we obtain
\beq
\lambda_{f_0}=(0.19\pm0.02)\GeV^2.
\lb{lam}
\enq
We can proceed and evaluate
the $f_0(980)$ mass from the above sum rule, taking the
derivative of Eq.~(\ref{2f0}) with respect to $M^{-2}$ and dividing the 
resulting sum rule by Eq.~(\ref{2f0}). For $u_{0}=1.6~\GeV^2$ and the 
mixing angle $\alpha=37^0$, the mass is in a very good agreement 
with the experimental number and is very stable, as a function of the
Borel mass. In the considered Borel range we 
obtain: $m_{f_0}=(0.98\pm0.01)\GeV$.
 For smaller (bigger) values of
$\alpha$ we get a bigger (smaller) mass. The best agreement with
the experimental result is obtained for $\alpha\sim 37^0$. 

The coupling given in Eq.~(\ref{lam}) will be used in the analysis of
the semileptonic decays of $D_s$ and $D$ into $f_0(980)$. The
$D_I\to f_0(980)\ell\nu_\ell$ form factors are defined through the
matrix elements
\beq
\lag f_0(\pli)|\bar{s}\ga_\mu(1-\ga_5)c|D_s(p)\rag=\cos(\alpha)\left(
f_+^{D_s}(t)(p+\pli)_\mu+f_-^{D_s}(t)q_\mu\right),
\label{f+ds}
\enq
and
\beq
\lag f_0(\pli)|\bar{d}\ga_\mu(1-\ga_5)c|D(p)\rag={\sin(\alpha)
\over\sqrt{2}}\left(
f_+^{D}(t)(p+\pli)_\mu+f_-^{D}(t)q_\mu\right),
\label{f+d}
\enq
with $t=q^2$ and $q=p-\pli$. Since in the decay rate the form factor $f_-(t)$
is multiplied by the difference of the lepton masses, its contribution
is negligible for both $e$ and $\mu$ decays. Therefore,
the differential semileptonic decay rates are given by
\beq
\frac{d\Ga(t)}{dt} =\cos^2(\alpha)
\frac{G_F^2|V_{cs}|^2}{192\pi^3m_{D_s}^3} \la^{3/2}(m_{D_s}^2,m_{f_0}^2,t)
(f_{+}^{D_s}(t))^2\;,
\lb{gads}
\enq
for $D_s^+\to f_0(980)\ell^+\nu_\ell$, and
\beq
\frac{d\Ga(t)}{dt} ={\sin^2(\alpha)\over2}
\frac{G_F^2|V_{cd}|^2}{192\pi^3m_{D}^3} \la^{3/2}(m_{D}^2,m_{f_0}^2,t)
(f_{+}^{D}(t))^2\;,
\lb{gad}
\enq
for $D^+\to f_0(980)\ell^+\nu_\ell$. In Eqs.~(\ref{gads}) and (\ref{gad})
$\la(x,y,z)=x^2+y^2+z^2-2xy-2xz-2yz$,
$G_F$ is the Fermi coupling constant and $V_{cs}$ and $V_{cd}$ are the 
Cabibbo-Kobayashi-Mashawa transition elements.

Using the QCD sum rule technique \cite{svz}, the form factors in 
Eqs.~(\ref{f+ds}) and (\ref{f+d})  can be evaluated from the time ordered 
product of the interpolating fields for $D_I$ and $f_0$, and 
the weak current $j^W_\mu=\bar{q_I}\gamma_\mu(1-\gamma_5) c$ (where
$q_I$ is $s$ or $d$ for $D_I$ being $D_s$ or $D$ respectively):
\beq
T_{\mu}(p,\pli) = i^2\int d^4x d^4y\, \lag0|{\rm T}[ j_s(x) j^W_\mu(y)
j_{D_I}^\dagger(0)]|0\rag e^{i(\pli.x+q.y)}\;,
\lb{3point}
\enq
where the $D_I^+$ meson in the initial state is interpolated by the 
pseudoscalar current
\beq
j_{D_I}(x) = \bar{q_I}(x)i\ga_5 c(x)\;,
\lb{intD}
\enq
and the  $f_0(980)$ is interpolated by the scalar current given in
Eq.~(\ref{jf0}). 

In order to evaluate the phenomenological side
we insert intermediate states for $D_I$ and $f_0$, we use the definitions 
in Eqs.~(\ref{f+ds}) and (\ref{f+d}), and obtain the following relations
\beqa
&&T_{\mu}^{phen} (p,\pli)={m_{D_s}^2f_{D_s}\over m_c+m_s}\lambda_{f_0}
\cos(\alpha){f_{+}^{D_s}(t)(p+
\pli)_\mu+f_{-}^{D_s}(t)q_\mu\over (m_{D_s}^2-p^2)(m_{f_0}^2-{\pli}^2)}
\nn \\
&&+\mbox{ contributions of higher resonances}\;,
\lb{phends}
\enqa
for $D_s^+\to f_0(980)$ and
\beqa
&&T_{\mu}^{phen} (p,\pli)={m_{D}^2f_{D}\over m_c}\lambda_{f_0}
{\sin(\alpha)\over\sqrt{2}}{f_{+}^{D}(t)(p+
\pli)_\mu+f_{-}^{D}(t)q_\mu\over (m_{D}^2-p^2)(m_{f_0}^2-{\pli}^2)}
\nn \\
&&+\mbox{ contributions of higher resonances}\;,
\lb{phend}
\enqa
for $D^+\to f_0(980)$.

In the above equations we have used the standard definition of the 
couplings of $D_I$ with the corresponding current:
\beq
\lag 0 | j_{D_I}|D_I\rag ={m_{D_I}^2f_{D_I}\over m_c+m_{q_I}}\;.
\lb{ds}
\enq

The three-point function Eq.(\ref{3point}) can be
evaluated by perturbative QCD if the external momenta are in the deep 
Euclidean region
\beq
p\ll (m_c+m_s)^2,~~~ {\pli}^2 \ll 4m_s^2, ~~~ t \ll (m_c+m_s)^2\;.
\lb{cond} 
\enq
In order to approach the not-so-deep-Euclidean region and to 
get more information on the nearest physical singularities, 
nonperturbative power corrections are added to the perturbative contribution.
In practice, only the first few condensates contribute significantly, the
most important ones being the 3-dimension, quark condensate,
and the 5-dimension, mixed (quark-gluon) condensate.
For the invariant structure, $(p+\pli)_\mu$, we can write
\beqa
T^{theor}(p^2,{\pli}^2,t)&=& {-1\over4\pi^2}\int_{(m_c+m_{q_I})^2}^\infty ds
\int_0^\infty du\,
\frac{\rho_+(s,u,t)}{(s-p^2)(u-{\pli}^2)}
\nn\\
&+& T^{D=3}_{+}\lag\bar{q_I}q_I\rag+ T^{D=5}_{+}\lag\bar{q_I}g_s\si.Gq_I\rag+
\cdots\;.
\lb{power}
\enqa
The perturbative contribution is contained in the double discontinuity
$\rho_{+}$.

In order to suppress the condensates of higher dimension and at the same time
reduce the influence of higher resonances, the series in Eq.~(\rf{power}) is
double Borel improved. Furthermore, we make the usual assumption that the 
contributions of  higher
resonances are well approximated by the perturbative expression
with appropriate continuum thresholds $s_{0}$ and $u_{0}$.
By equating the Borel transforms of the phenomenological expression
for the  $(p+\pli)_\mu$ invariant structure in Eqs.(\rf{phends})
and (\ref{phend}), and that of the ``theoretical expression'', 
Eq.~(\rf{power}), we obtain the sum rule for the form factor $f_+^{D_I}(t)$.
The sum rule for $f_+^{D_s}(t)$ (at the order $m_s$) is given by:
\beqa
&&
f_+^{D_s}(t)~C~e^{-m_{D_s}^2/M^2}e^{-m_{f_0}^2/\mli}
={-1\over4\pi^2}\int_{(m_c+ms)^2}^{s_0} ds 
\int_0^{u_0}du\left[e^{-s/M^2}e^{-u/\mli}\rho_+(s,u,t)\right]
\nonumber\\
&+&{\over2}e^{-m_c^2/M^2}\left[-m_c+2m_s+{m_c^2m_s\over2M^2}
\right]+m_0^2\lag\bar{s}s\rag e^{-m_c^2/M^2}\left[{m_c^2(m_c-m_s)\over8M^4}
-{2m_c-m_s\over6M^2}\right.\nonumber\\
&+&\left.{m_c^2(4m_c-3m_s)-2t(m_c-m_s)\over24M^2\mli}-{m_c-2m_s\over6\mli}
+{m_c^2m_s-2t(m_c-m_s)\over24M^2\mli}
\right],
\lb{f+}
\enqa
where $C={m_{D_s}^2f_{D_s}\over m_c+m_s}\lambda_{f_0} $
and
\beqa
\rho_+(s,u,t)&=&{3\over\la^{3/2}(s,u,t)}\left\{u\left[2m_cm_s(2m_c^2-s-t+u)
+m_c^2(s-t+u)+s(-s+t+u)\right]\right.\nonumber\\
&-&\left.(2m_c^2-s-t+u)(su+m_cm_s(s-t+u))\right\}~\Theta(s-s_M),
\enqa
with $s_M=m_c^2+{m_c^2u\over m_c^2-t}$. The sum rule for $f_+^D(t)$ can be 
obtained from Eq.~(\ref{f+}) by just neglecting the $m_s$ terms, changing
$\lag\bar{s}s\rag$ by $\lag\bar{q}q\rag$ and $D_s$ by $D$.

The decay constant $f_{D_I}$ defined in Eq.~(\rf{ds}), and appearing in the
constant $C$, can also be determined  by sum rules
obtained from the appropriate two-point functions \cite{phi,ck00}.

The value of the charm quark and meson masses are:
$m_c=1.3\,\GeV$, $m_{f_0}=\,0.98\GeV$, 
$m_{D_s}=1.97\,\GeV$ and $m_D=1.87\,\GeV$. For the $D_s$ and $D$
decay constants we use $f_{D_s}=(0.22\pm0.02)\,\GeV$ 
and $f_D=(0.17\pm0.02)~\GeV$ \cite{phi,ck00,lat2}.
For the continuum thresholds we take the values discussed in 
refs.~\cite{faz,phi,ck00}:
$s_{0}=(7.7\pm1.1)~\GeV^2$ for $D_s$, $s_0=(6.0\pm0.2)~\GeV^2$ for $D$
and $u_0=(1.6\pm0.1)~\GeV$.

\begin{figure} \label{fig2}
\centerline{\psfig{figure=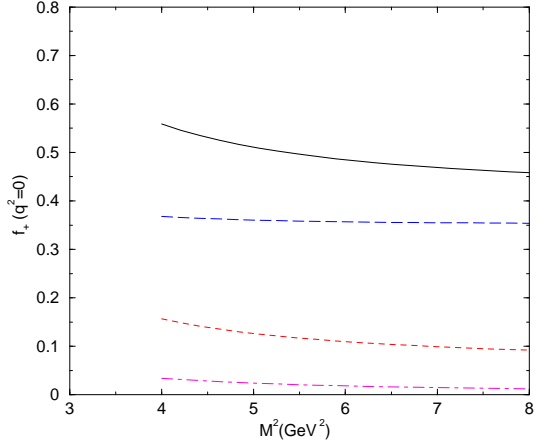,height=60mm,width=70mm,angle=0}
}
\caption{Various contributions to the OPE of the form factor $f_+^{D_s}(t)$,
at zero momentum transfer, as a
function of the Borel parameter $M^2$. 
 Solid curve: total contribution; long-dashed: perturbative; 
dashed: quark condensate; dot-dashed mixed condensate contribution.} 
\end{figure}

We evaluate our sum rules in 
the range $4.0\leq M^2\leq8.0\GeV^2$, at a fixed ratio 
$M^2/\mli = (m_{D_I}^2-m_c^2)/m_{f_0}^2$, which is compatible with the
Borel ranges used for the two-point functions in refs.~\cite{faz,phi}.
In Fig.~2 we show the different  contributions to the form
factor $f_+^{D_s}$ at zero momentum transfer, from the sum rule
in Eq.~(\ref{f+}), as a function of the Borel variable $M^2$. 
We see that the 
perturbative contribution is the largest one, and that the 
mixed condensate contribution is negligible. A similar behavior is also
obtained  for $f_+^{D}(t)$.
Varying the continuum
thresholds $s_0$ and $u_0$, and the couplings $\lambda_{f_0}$,
$f_{D_s}$ and $f_D$ in the ranges given 
above,  we get for the form factors at $t=0$:
\beq
\begin{array}{c}
f_+^{D_I}(0)=\end{array}
\left\{
\begin{array}{c}
0.50\pm0.13~\mbox{ for } D_s\\
0.53\pm0.15~ \mbox{ for } D
\end{array}
\right.
\lb{fornum}
\enq
Our values for $f_+^{D_s}(0)$ are compatible
with the values found in refs.~\cite{cheng,detc}, where the 
nonleptonic decay of the $D_s^+$ meson were studied.

The $t$ dependence of the form factors evaluated at $M^2=7\GeV^2$
in the range  $-0.5\leq t \leq 0 \; \GeV^2 $ 
can be fitted by a linear expression 
\beq
f_+^{D_I}(t)=f_+^{D_I}(0)+At\;.
\enq
This expression is consistent, in the considered range of momentum transfer,
with a monopole expression 
$f_+^{D_I}(t)={f_+^{D_I}(0)\over 1-{t\over M_P^2}}$,
with the mass of the pole $M_P=(1.70\pm0.05)~\GeV$ for $D_s$ and
$M_P=(1.95\pm0.05)~\GeV$ for $D$. It is interesting to notice
that $M_P$ is compatible with the values found for the decays 
$D_s\rightarrow\eta$ \cite{cofa} and $D\to\kappa$ \cite{sca}.

In the  limits of the variables and the continuum thresholds discussed
above we obtain for the semileptonic decay widths
\beqa
\Gamma(D_s^+\to f_0(980)\ell^+\nu_\ell)=\cos^2(\alpha) (8.1 \pm 4.1)~\GeV\;, 
\nn\\
\Gamma(D^+\to f_0(980)\ell^+\nu_\ell)=\sin^2(\alpha) (1.5 \pm 0.8)~\GeV\; , 
\label{dre}
\enqa
where we have used $V_{cs}=0.975$ and $V_{cd}=0.22$. The ratio
between the two decay widths given in Eq.~(\ref{dre}) gives us a direct
information about the mixing angle:
\beq
{\Gamma(D_s^+\to f_0(980)\ell^+\nu_\ell)\over
\Gamma(D^+\to f_0(980)\ell^+\nu_\ell)}={56\pm2\over\tan^2(\alpha)}\,.
\lb{rat}
\enq
For $\alpha\sim37^0$, from Eq.~(\ref{rat}) we can conclude that the
semileptonic decay width of $D_s$ into $f_0(980)$ would be around one 
hundred times larger than the semileptonic decay width of $D$ into 
$f_0(980)$. Since in the semileptonic decays there are no complications
due to strong interactions, we believe that an experimental measurement
of the ratio in Eq.~(\ref{rat}) is the cleanest way to evaluate
the mixing angle $\alpha$. 

It is interesting to notice that if the current in Eq.~(\ref{jf0}) were used
in ref.~\cite{phi} for $f_0(980)$, instead of a pure $\bar{s}s$ current,
the ratio calculated there would change to
\beq
R={\Ga(D_s^+\rightarrow f_0(980)\pi^+)\over \Ga(D_s^+\rightarrow 
\phi\pi^+)}=\cos^2(\alpha)(0.44\pm0.18)\;.
\label{fin}
\enq
Therefore, for $\alpha\sim37^0$, which is compatible with the mixing angle
found in refs.~\cite{cheng,adn}, the above ratio would be reduced to
$R=(0.28\pm0.11)$ in agreement with the experimental result
$R^{exp}=(0.210\pm0.069)$ \cite{E791}. 

At this point we might conclude by saying that $f_0(980)$ is well described
by the relation in Eq.~(\ref{f0stru}), with the mixing angle being
$\alpha\sim 35^0$, and by making the prediction in Eq.(\ref{rat}).
However, there are two experimental facts which do not fit in this
picture and that might even be in contradiction with Eq.(\ref{f0stru}):
\begin{itemize} \item A mixing angle in $f_0(980)$ also implies
that $\sigma$ would have a strange component (see Eq.(\ref{f0stru})). 
Therefore, the decay $D_s^+\to\sigma\pi^+$ should also occur. As a matter
of fact, the ratio ${\Ga(D_s^+\rightarrow \sigma\pi^+)/\Ga(D_s^+
\rightarrow f_0(980)\pi^+)}$ is proportional to $\tan^2(\alpha)$. However,
the E791 Collaboration  did not observe any contribution
from $D_s^+\rightarrow \sigma\pi^+$ to the $D_s^+\rightarrow 
\pi^+\pi^+\pi^-$ decay \cite{E791} pointing towards $\alpha\sim0$.
\item In the framework of generalized factorization, the amplitude for
the $D^+\rightarrow f_0(980)\pi^+$ decay is given by (neglecting the 
annihilation term)
\beq
A(D^+\to f_0(980)\pi^+)={G_F\over\sqrt{2}}V_{cd}V^*_{ud}
\left(c_1(\mu)+{c_2(\mu)\over 3}
\right)f_\pi(m_D^2-m_{f_0}^2){\sin(\alpha)\over\sqrt{2}}f_+^D(0)\;,
\label{df0}
\enq
since $\lag0|V_\mu|f_0\rag=0$ due to charge conjugation invariance
and conservation of vector current. In Eq.(\ref{df0}) $c_i$ is the Wilson
coefficient entering the effective weak Hamiltonian.
Therefore, using the results in Eq.~(\ref{fornum}) we get
\beq
{\Gamma(D_s^+\to f_0(980)\pi^+)\over
\Gamma(D^+\to f_0(980)\pi^+)}={46\pm2\over\tan^2(\alpha)}\,.
\lb{rat2}
\enq

From the  E791 Collaboration, the experimental value
of this quantity is \cite{E791}:
\beq
\left({\Gamma(D_s^+\to f_0(980)\pi^+)\over
\Gamma(D^+\to f_0(980)\pi^+)}\right)^{exp}=13.3\pm0.4\,,
\lb{raexp}
\enq
which would lead to $\alpha\sim62^0$. 
\end{itemize} 
 
 It is important to remember
that in the analysis of the nonleptonic decays we are using the 
factorization approximation. Therefore, the apparent inconsistency
between the data and the mixing angle can still be due to this approximation.
This is why the measurement of the ratio in
Eq.~(\ref{rat}) would bring important information about this
puzzle. If the light scalars could not be seen in the semileptonic decays, 
this would clearly indicate that their structure is more complicated that
simple quark-antiquark states. One possibility is that they are
four-quark states, as suggested in refs.~\cite{cloto,jaffe,lat}.

To summarize,
we have presented a QCD sum rule study of the $D_s^+$ and $D$ semileptonic 
decays to $f_0(980)$, considered as a mixture of the scalars $\bar{s}s$ and 
$\bar{u}u+\bar{d}d$.
 We have evaluated the $t$ dependence of the form factors
$f_+^{D_I}(t)$ in the region $-0.5\leq t\leq0\GeV^2$.
The $t$ dependence of the form factors could be fitted by a linear form
compatible, in the studied range, with a monopole form,
and extrapolated to the full kinematical region. 

The form factors were used to evaluate the decay widths of
the decays $D_s^+\to f_0(980)\ell^+\nu_\ell$ and $D^+\to f_0(980)\ell^+
\bar\nu_{\ell}$ as a function of the mixing angle. Experimental data
about these decays would provide a direct estimate of the mixing angle.
Using the same form factors to evaluate nonleptonic decays in
the framework of the factorization approximation, it was not possible to 
explain all available experimental data with a fixed mixing angle.
However, data seem to suggest that there is a sizable
non strange component in the $f_0(980)$ meson.

\vspace{1cm}
 
\underline{Acknowledgments}: 
This work has been supported by CNPq and FAPESP (Brazil).

\end{document}